\documentclass{osa-article}
\usepackage{graphicx}

\journal{osajournal}

\articletype{Research Article}

\begin{document}

\title{Subwavelength effects near a dielectric microcylinder illuminated by a diffraction-free beam}

\author{R.~Heydarian,\authormark{1,*} V.~Klimov,\authormark{2} and C.~Simovski\authormark{1,3}}

\address{\authormark{1}Department of Electronics and Nano-Engineering, Aalto University, P.O. Box 15500, FI-00076 Aalto, Finland\\
\authormark{2}Department of Optics, Lebedev Physical Institute, Russian Academy of Science, 53, Leninsky Prospekt, 119991, Moscow, Russia\\
\authormark{3}Faculty of Physics and Engineering, ITMO University, 199034, Birzhevaya line 16, Saint-Petersburg, Russia}

\email{\authormark{*}reza.heydarian@aalto.fi} 



\begin{abstract}
Generation of a photonic nanojet by a linearly polarized wave beam or a plane wave impinging a dielectric microcylinder 
implies partial conversion of the propagating waves into the evanescent ones. This conversion is manifested in nanojet waist of 
subwavelength effective width. However, this known near-field effect is relatively weak. 
We theoretically show that the incidence of a wave beam formed by two plane waves enables 
much stronger near-field effects: a deeply subwavelength focusing of the incident beam and a strong enhancement of the electric field on the whole cylinder surface and near it. The domination of the evanescent waves in the vicinity of the cylinder results from the destructive interference of the propagating spatial harmonics of the scattered field dictated by the incident wave beam. 
\end{abstract}

\section{Introduction and problem formulation}

\begin{figure}[htbp]
\centering

\includegraphics[width=8cm]{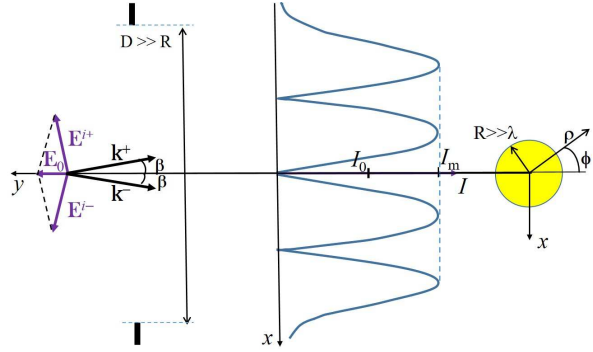}

\caption{Planar problem: a cosine beam impinges a dielectric microcylinder. Cartesian and cylindrical coordinates systems are shown, 
important notations are depicted.}
\label{Pic1}
\end{figure}

In this paper we consider the near-field effects which arise when a 2D diffraction-free beam impinges a dielectric microcylinder with optically large radius $kR\gg\pi$ (here and below $k=2\pi/\lambda$ is the wave number of free space). Our incident beam is a 2D analogue of the radially polarized Mathieu beam \cite{radial} and is called in the theory of diffraction the cosine wave beam \cite{cosine}. In practice, such a beam results from the transmission of two plane waves though a large diaphragm $D\gg\lambda$. These waves have the same frequency and opposite phases and their wave vectors should form a sharp angle $2\beta$ as it is depicted in Fig.~\ref{Pic1}. From the aperture $D$ till the distances exceeding $D$, the cosine beam experiences the Abbe diffraction only in its tails. In the paraxial domain of the beam only the interference of two plane waves is observed. 
For our purposes, the presence of the diaphragm is not relevant. In our simulations, we saw that the power flux distant from the cylinder lateral edges
$x=\pm R$ by $\Delta x> \lambda$ do not feel the presence of the cylinder i.e. the relevant part of the incident beam is $-R-\lambda<x<R+\lambda$. 
In this part the diaphragm has no any impact, and in our simulations reported below the cosine beam represents simply two plane waves with the angle $2\beta$ between their wave vector ${\bf k}^{\pm}$. 

If this angle is small enough ($2\beta<\pi/kR$) and the beam is TM-polarized the $x$-component of the incident electric field and the incident magnetic field grows versus $|x|$ from zero almost linearly over the interval $|x|=[0,R]$. In this case the period of the incident beam intensity versus $x$ is optically large, and the maximal intensity (square of the electric field amplitude) of the incident beam $I_i(x)=I_m$ holds when $|x|> R$. In this paper, we will show that this structure of the incident beam enables very strong near-field effects in the area behind the cylinder and on its surface. 

Since the cylinder practically feels the incident beam which only within the interval $x=[-R,R]$ 
it is reasonable to normalize the electric intensity $I$ of the total field     
to the intensity $I_0$ of the incident beam averaged over this relevant interval. The value $I_0$ is marked in Fig.~\ref{Pic1}.
The subwavelength concentration of the electric field we aim to obtain obviously implies high values of the 
local intensity enhancement $I/I_0$.   

Resonant enhancement of the electric intensity at some points inside a dielectric microcylinder represents nothing new. This effect was described in several books, see e.g. in \cite{Sergei}. Below, we will show similar features in free space, not only on the surface of the microcylinder, but also at a distance behind it. In this area an only near-field effect known for a dielectric microcylinder is the waist of the so-called photonic nanojet \cite{PNJ}. A photonic  nanojet is a wave beam arising on the back side of a dielectric particle (a microcylinder or a microsphere) illuminated by a plane wave. On the back surface of the microparticle and at a small distance behind it this beam has a waist of subwavelength effective width \cite{PNJ,Itagi,Taflove,Taflove1}. In the domain of this waist, the electric field comprises a noticeable evanescent-wave component, a longitudinal component of the electric field arises there and the electric intensity enhances several times compared to that of the incident wave \cite{Itagi,Taflove1}. However, this known near-field effect is quite weak. For a microsphere, the waist effective width calculated at the rear sphere surface is within the range $(0.29-0.35)\lambda$ \cite{PNJ,Itagi,Taflove,Taflove1}. For a microcylinder, this width is $(0.4-0.5)\lambda$, whereas the local enhancement of the electric intensity inside the waist does not exceed $4$ \cite{Taflove1}. At a distance of about $2\lambda$ from the microcylinder the Abbe diffraction turns this waist into a usual wave beam of Gaussian type. It is clear, that the generation of the photonic nanoject by a plane wave is such the combination of near-field and far-field effects, in which the latter effects dominate.    

Our case is opposite because the problem symmetry prohibits the generation of the photonic nanojet. On the axis $y$ the total electric field is longitudinal ($E_{\rho}(\phi=0)=E_y\ne 0$, $E_{\phi}=0$) and the magnetic field nullifies. Suppressing this way the propagating part of the total field spatial spectrum in the paraxial region our wave beam offers the enhancement to the evanescent components of this spectrum. This point will be additionally discussed below. We will show that the rear point of the cylinder $(x=0, y=R)$, or, in cylindrical coordinates $(\phi=0,\rho=R)$, may be a center of a deeply subwavelength focal spot. The magnitude of the electric field $E$ at this point may be enhanced drastically compared the incident beam electric field magnitude within the relevant interval $-R<x<R$. This regime is granted by the incident wave beam symmetry as it is shown in Fig.~\ref{Pic1} and by the proper choice of the wave tilt angle $\beta$ (it should be smaller than $\pi/2kR$).    

Let us explain how we came to the last idea. The initial motivation of this study was related to the nanoimaging granted by a simple dielectric microsphere. This label-free subwavelength image in presence of the microsphere was experimentally observed in work \cite{Hong} and since that time 
many researchers have studied this approach to the nanoimaging experimentally. However, the physics of this effect have not been fully understood up to now. Some particular experiments were explained via the resonances of the whispering gallery, via multipole Mie resonances of a spherical cavity, and for plasmonic substrates and plasmonic objects -- via the  excitation of surface plasmons in the nanogap between the sphere and the substrate \cite{Yang,Lecler,Zhou,Maslov,Cang}. However, after these works many available experimental results left unexplained or were explained in a wrong way (see more details in \cite{Astratov1}).

In our recent work \cite{Reza} we suggested two additional physical mechanisms which explain the far-field imaging of a glass microsphere 
for those cases which were not explained in the precedent works. Both these mechanisms are related with the radial polarization of the scattering object with respect to the sphere. Since the near-field coupling of a dipole radially polarized with respect to the sphere results in the concentration of the 
near electric field in a subwavelength domain adjacent to the dipole, there is a small spot of the induced polarization that serves a source for the so-called imaging beam created by the microsphere. 

One of two assumed physical mechanisms corresponded to the particular cases when  
the eigenmodes in the microsphere are uniformly and weakly excited, and form inside the sphere a quasi-continuum. 
In this case, to our assumption, the imaging beam is formed in accordance to the geometrical optics. This speculation results for a glass microsphere ($n=1.4-1.7$) in a beam with radial polarization (with respect to the propagation axis $y$) and azimuthally uniform intensity distribution. For $n\approx 1.4$ the intensity profile of the beam in the geometrical optics approximation is close to that of a non-divergent and polarized radially (with respect to the optical axis) Mathieu beam. This beam is diffraction-free and therefore a usual microscope can focus it into a subwavelength spot \cite{radial}. The subwavelength spot imaging a point dipole automatically implies the subwavelength resolution.        

In \cite{Reza} we have not confirmed this assumption by numerical simulations even for the 2D case because it is not realistic to simulate simultaneously a microsphere and an aberration-free objective lens which is a macroscopic object. Usually, to analyze the focusing of the imaging beam by a microscope one calculates the so-called point-spread function (see e.g. in \cite{Astratov1}). In the present case, it is difficult because  the polarization of our imaging beam is radial and the usual scalar formalism is not applicable. Therefore, we preferred to search an aberration-free focusing functionality of a microsphere impinged by such the Mathieu beam. It would allow us to simulate the whole system using one solver. 

On the first stage of this numerical study, we impinged a glass microsphere or a microcylinder (the physics of the nanoimaging in the 2D problem is basically the same, and the replacement of a sphere by a cylinder saves a lot of the computational time) by a diffraction-free beam with manually determined parameters. The goal of this study was to find the conditions of the subwavelength focusing of such the beam by a dielectric microparticle. On the second stage, we aimed to obtain the needed wave beam using another microsphere (microcylinder) coupled with a normally (with respect to its surface)  polarized dipole source. This beam has been already obtained in our simulations where a ultimately subwavelength dipole antenna excited a glass microcylinder being sandwiched between it and a substrate. Its intensity distribution is nearly the same as in the cosine beam corresponding to $D=4R$, $\beta=\pi/2kR$. Using the imaging sphere (forming this beam) of larger radius than that of the lens sphere (focusing this beam) on 
the last stage of our study we planned to simulate the whole system i.e. to show the subwavelength focal spot imaging our dipole. 
We expected for the system of two spheres to obtain the focal spot of the width close to $0.2\lambda$ corresponding to the right focusing of the radially polarized Mathieu beam in work \cite{radial}. For the system of two cylinders we expected a larger focal spot. 

However, on the first stage of this study we obtained unexpected results which seem to be more important than the whole planned study. 
Really, our planned study only could confirm a hypothesis suggested in \cite{Reza} so that to explain an already known effect. Meanwhile, on the first stage we have revealed unexpected effects. Therefore, we postpone reporting on the second and third stages of the initially planned research and in the present paper describe these new effects: the strongly subwavelength concentration of the electric field behind the rear edge of a glass microcylinder.
It corresponds to the focal spot $\lambda/6$ and very high local field intensity enhancement -- several dozens. Moreover, we have revealed a high field enhancement in an optically large area -- on the whole surface of the cylinder. Below we present and discuss these results.

\section{Simulations: results and discussion}

\subsection{Subwavelength focusing}

For reliability of our calculations we use two methods: commercial solver COMSOL Multiphysics and the well-known analytical solution of the 2D diffraction problem for a cylinder impinged by a plane wave (see e.g. in \cite{Sergei}). For our incident beam of two plane waves with the symmetric polarization the series of cylindrical functions expressing this solution converge much faster than  for a single plane wave. Therefore, for optically large cylinders $kR\gg \pi$ there is no need to use high-frequency asymptotics, as well as for a large dielectric sphere illuminated by Bessel beams of nonzero order studied in work \cite{Klimov}. Our incident beam propagating along $y$ has an only $z$-component of the magnetic field (we denote its magnitude by $A$) and the incident electric field on the axis $y$ is longitudinal:
\begin{equation}
H^i=H^i_z=A \sin (k\rho\sin\beta\sin\phi) e^{-jk\rho\cos\beta \cos\phi},\quad E^i(\phi=0)=E^i_{\rho}=E_0 e^{-jk\rho\cos\beta},
\end{equation}
where $E_0= -jA\sqrt{\mu_0/\varepsilon_0}\sin\beta$. For the mean electric intensity of the incident beam we have $I_0=E_0^2\sin^2(kR\sin\beta)/\sin^2\beta$. 
In the reported simulations $\beta=0.01$ and $\pi\ll kR<\pi/\beta$ i.e. an optically large cylinder is fully located in between two adjacent maxima of the incident beam. It implies that $E_0^2\ll I_0$.

If $kR\gg \pi$ any incident wave beam with any wave number $k$ for any refractive index $n$ of the cylinder efficiently excites in it a number of TM-polarized eigenmodes whose amplitudes are proportional to that of the incident beam. These eigenmodes (leaky as in any other dielectric cavity) resonate for given $n$ at specific values of $kR$ and for given $kR$ at specific $n$. Every mode has a number of such resonances. We are looking for the regimes of the incident beam subwavelength focusing at the point $(\phi=0, \rho=R)$. Subwavelength concentration of the electric energy implies the strong enhancement of the electric intensity $I=E^2_{\rho}$ at this point. What can cause this enhancement? Definitely, it may result from the resonances of the eigemodes.

\begin{figure}[htbp]
\centering
\includegraphics[width=8cm]{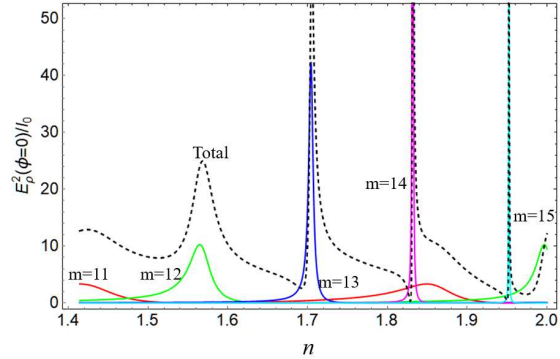}
\caption{Total (dashed line) and partial electric intensities of the eigenmodes ($m=11$ is red, $m=12$ is green, $m=13$ is blue, $m=14$ is magenta, $m=15$ is cyan) calculated at the point $(\phi=0, y=R)$ for the case $kR=10$ versus the refractive index of the cylinder. All intensities are normalized to $I_0$.}
\label{Pic2}
\end{figure}

Therefore, we start from the study of the modal electric field excited by our beam at the point of our main interest $(\phi=0,\rho=R)$ varying $n$ and $kR$. Using the analytical solution, we calculated the partial intensities of the TM modes $E_{m\rho}^2$ excited by our beam at this point. These partial intensities, normalized to $I_0$, are shown in Fig.~\ref{Pic2} as the functions of $n$ for the case $kR=10$. In this case, the modes of the orders $m=11-15$ are resonant within the interval $n=[1.4, 2.0]$  where some resonances are repeating. In the similar plots, calculated for larger $kR$, the density of the resonances of $E_{m\rho}^2$ over the axis $n$ is higher because for the thicker cylinder the higher modes also become resonant. Similarly, the set of resonant regimes arises for given $n$ versus normalized frequency $kR$ and, similarly, the increase of $n$ increases the density of the resonances on the axis $kR$.    

Total intensity $I(0,R)$ results from the interference of these modes with one another and with the incident beam. However, 
as we can see in Fig.~\ref{Pic2} $E_{m\rho}^2$ for $m=11-15$ exceed $I_0$ for any $n$ because the resonant ranges of $n$
corresponding to different $m$ intersect. Roughly speaking the whole range $n=1.4-2$ is more or less resonant. 
Since $I_0\gg E_0^2$ this fact means that $E_{m\rho}^2(0,R)$ exceed the incident field intensity at the rear edge of the cylinder drastically. In other words, in the value $I(0,R)$ the contribution $E_0^2$ of the incident beam is negligibly small, and $I(0,R)$ is determined only by the interference of the excited eigenmodes. Studying the phases of these modes we have found that for $kR=10$ the intensity $I(0,R)$ has local maxima versus $n$ which do not exactly coincide with the resonant values of $n$ seen in Fig.~\ref{Pic2}.

\begin{figure}[htbp]
\centering
\includegraphics[width=12cm]{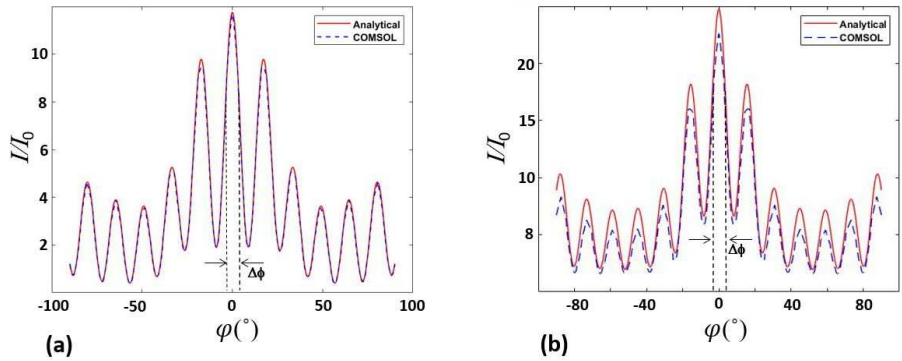}
\caption{Electric intensity enhancement on the back side of the cylinder for $n=1.40$ (a) and $n=1.57$ (b).
}
\label{Pic3}
\end{figure}

In Fig.~\ref{Pic3} we present the plots of $I(\phi,R)/I_0$ calculated on the back side of the cylinder surface for (a) $n=1.40$ and (b) $n=1.57$. 
Analytical solution is in a very good agreement with the result of COMSOL in both cases. In Fig.~\ref{Pic3}(a) the angular width $\Delta\phi$ of the central intensity maximum determined in accordance to the Rayleigh criterion (0.7 of the maximum) is nearly $8.3^{\circ}$, that implies the linear width of the focal spot $R\Delta\phi\approx 0.24\lambda$. In Fig.~\ref{Pic3}(b) $\Delta\phi\approx 7.2^{\circ}$ and the focal spot is narrower: $R\Delta\phi\approx 0.20\lambda$. In this case, the refractive index is close to the resonant values for the modes $m=12$ and $m=14$, and this is the reason why the focal spot for $n=1.57$ is  more subwavelength than it is for $n=1.4$. Also, the local intensity enhancement at the focal point for $n=1.57$ is as high as $22-24$ ($24$ results from the analytical model and $22$ from COMSOL), whereas for $n=1.4$ we have $I/I_0\approx 11.5$ in both COMSOL and analytical solution. We have not found in the available literature, local intensity enhancement  due to the presence of any dielectric cylinder of micron or submicron radius exceeding $4$ for a point located in free space. It is clear that this effect results 
from a specific excitation of our cylinder.

\begin{figure}[htbp]
\centering
\includegraphics[width=11cm]{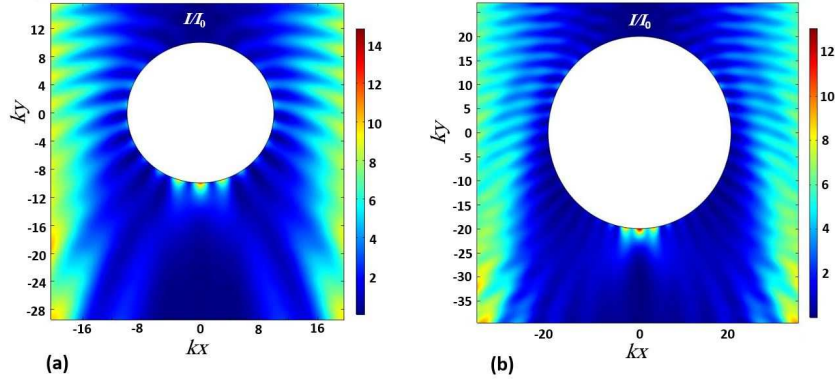}
\caption{Color map of the normalized intensity outside the cylinder when $kR=10,n=1.4$ (a) and the similar map when $kR=20,n=1.7$ (b).
Incidence from top. 
}
\label{Pic4}
\end{figure}

The color map of the normalized intensity $I/I_0$ around the cylinder for the case $kR=10,n=1.4$ is presented in Fig.~\ref{Pic4}(a). The central focal spot with subwavelength width is not a waist of a photonic nanojet (see above). Together with two lateral spots the focal spot represents the birthplace of two wave beams symmetrically tilted to the axis $y$. A set of similar beams leaks from the overlapping spatial maxima of the important modes ($m=11,\, m=14$) on both sides of the cylinder. These lateral beams cannot be visually distinguished in the regions $|x|\ge 20$ where they  become negligible on the background of the incident beam intensity maxima.     

The most important feature we can see in this color map is a clear spatial separation of the central maximum (focal spot) from two lateral maxima. Also, it is seen that the focal spot keeps the subwavelength width at sufficiently small distances $\Delta y$ from the rear edge of the cylinder. When $k \Delta y=1$ (i.e. in the plane $ky=-11$) the linear width of the focal spot in the plane $(x-z)$ is nearly equal $\Delta x =0.26\lambda$ i.e. is still subwavelength. For $n=1.57$ the width of the focal spot at the same distance from the cylinder is smaller: $\Delta x =0.22\lambda$. So, in  a plane located at a certain distance from the microcylinder the intensity distribution still forms a pronounced subwavelength spot.                

In Fig.~\ref{Pic4}(b) we present the similar color map for $kR=20$ and $n=1.7$. In this case, $n$ is close to the resonant value, i.e. the regime is similar to the case $kR=10$, $n=1.57$. In this regime, $\Delta\phi\approx 4.7^{\circ}$. It implies the width of the focal spot at the cylinder surface $R\Delta\phi\approx 0.27\lambda$. At the distance $\Delta y=1/k$ from the cylinder the width of the focal spot is $\Delta x\approx 0.29\lambda$ -- still subwavelength.

\begin{figure}[htbp]
\centering
\includegraphics[width=10cm]{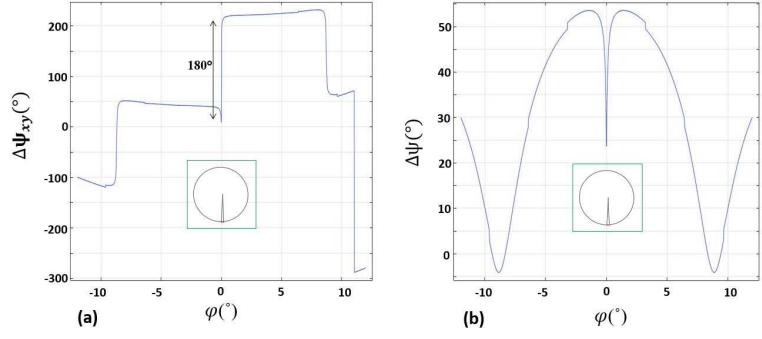}
\caption{(a) Distribution of the phase shift between $E_x$ and $E_y$ on the rear surface of the cylinder in a region covering three main local maxima of the electric field (this region is marked by red in the inset). (b) Phase shift between the electric and magnetic field vectors in the same area (b).
}
\label{Pic5}
\end{figure}

The subwavelength focusing of a wave beam definitely implies the conversion of propagating waves into evanescent waves.  
Fig.~\ref{Pic5} illustrates the domination of the evanescent waves on the rear surface of the cylinder $kR=10, n=1.4$ via the analysis of the phase shifts between the components of the electromagnetic field. Fig.~\ref{Pic5}(a) depicts the distribution of the phase shift $\Delta\Psi_{xy}$ between two Cartesian components of the electric field over the part of the microparticle surface marked by red in the inset. This area covers the central focal spot and two lateral ones in which
$E_x$ and $E_y$ have nearly the same magnitude. At the center of these spots the phase shift $\Delta\Psi_{xy}\equiv phase(E_x/E_y)$ passes through the values 
$\pm 90^{\circ}$ that means the circular polarization. In general, the electric polarization is elliptic. It is linear (and longitudinal) only on the  
axis $y$. Meanwhile the magnetic polarization keeps one-component polarization $H=H_z$ everywhere. 
The phase shift $\Delta\Psi$ between the elliptically polarized electric field vector and linearly polarized magnetic one exceeds $50^{\circ}$ within the central spot. 
These phase relations and corresponding polarization transformation effect clearly point out the domination of the eavnescent waves in the vicinity of the rear edge of our cylinder.

\begin{figure}[htbp]
\centering
\includegraphics[width=11cm]{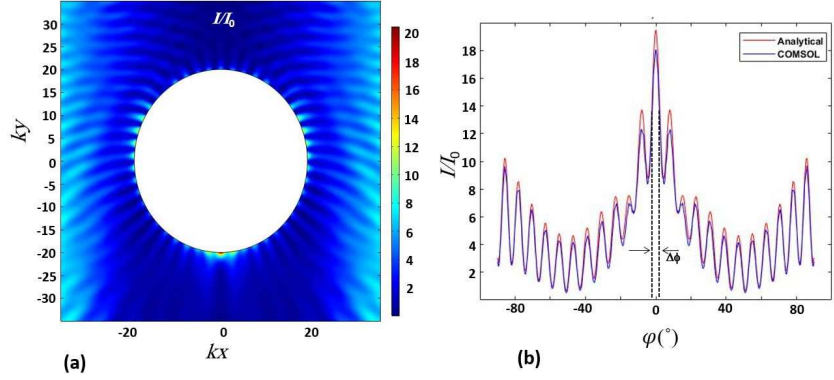}
\caption{Color map of the normalized intensity in the case of the tightest focusing, corresponding to $kR=20, n=1.61$ (a) and the plot of this intensity calculated on the back surface of the cylinder (b). Incidence from top.
}
\label{Pic6}
\end{figure}

To conclude this subsection let us report the best corresponding result. Varying $kR$ within the interval $10 \leq  kR\leq 20$ and the refractive index in the limits $1.4\leq n\leq 2$ we have found the case, when the linear width of the focal spot is minimal. This is the case when $kR=20$ and $n=1.610$. This refractive index is very close to the value $n=1.612$ corresponding to the superposition of two resonances mode resonances: $m=23$ and $m=27$. The local intensity enhancement for this case is depicted in Fig.~\ref{Pic6} in both color map and plot. The angular width of the focal  
spot on the surface is equal $3.1^{\circ}$, and the linear width is 
$R\Delta \phi \approx 0.15 \lambda$. At the distance $\Delta y=1/k$ i.e. in the plane $ky=-21$ the focal spot has the width $\Delta x \approx 0.16 \lambda$. We can see that in this case our analytical solution and COMSOL simulations are also in a good agreement. 

Thus, the possibility of subwavelength focusing of a diffraction-free beam by a microcylinder is proved. This focusing is accompanied by a strong enhancement of the electric field in the focal spot. The subwavelength focusing can be implemented with several values of $n$ for any $kR$ if the adopted condition $\pi\ll kR\ll \pi/\beta$ is respected. For given $n$ and $R$ this regime can be implemented (under this condition) for several values of $k$. Therefore, a cosine wave beam with a wide enough continuous frequency spectrum impinging a glass microcylinder completely located between its intensity maxima contains the spectral components which will be focused behind the cylinder into a subwavelength spot.

\subsection{Maximal enhancement of the surface-averaged electric intensity}

\begin{figure}[htbp]
\centering
\includegraphics[width=11cm]{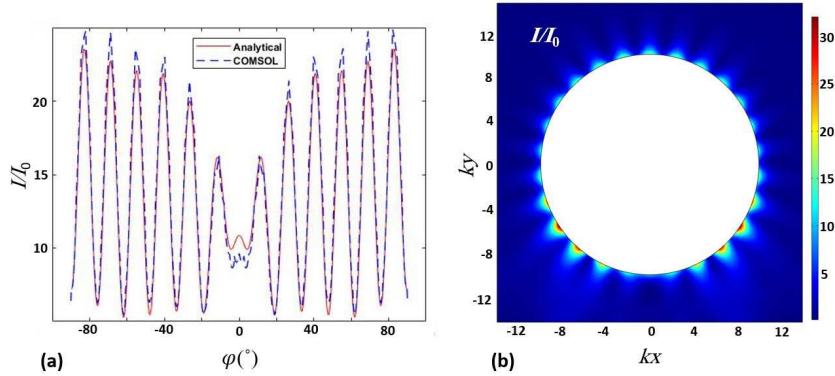}
\caption{Angular plot of the normalized intensity calculated on the back surface of a cylinder with $kR=10, n=1.7$ (a) and 
color map of intensity for the case (b).}
\label{Pic7}
\end{figure}

Above it was shown that the resonant magnitudes of the eigenmode electric field at the point $(\phi=0,\rho=R)$ are achieved for a number of 
the cylinder parameters. The other sets of the design parameters offer the resonant magnitudes of the electric field at the other points of the cylinder surface. It is even possible to find the regime, when the value $I/I_0$ averaged over $-90^{\circ}<\phi<90^{\circ}$ is maximized (for given $kR$ or for given $n$). For $kR=10$ the maximum of the mean value of $I/I_0$ (close to $15$) is achieved when $n=1.7$. This regime is illustrated by Fig.~\ref{Pic7}. We can see that the local maximum at the rear edge $(\phi=0,\rho=R)$ is relatively weak, whereas on the sides 
local intensity enhancement $I/I_0$ attains $23-24$. The effective width of all local maxima is weekly subwavelength (close to $0.4\lambda$). 
All intensity maxima except the axial one, in this regime can be treated as the waists of the photonic nanojets.   

Since this regime grants a significant enhancement of the electric intensity to the whole surface of the cylinder, it can be used for the enhancement of the fluorescence and Raman scattering of quantum emitters. For example, a long dielectric rod with optically substantial diameter can be covered by fluorescent molecules and impinged by a cosine beam. This technical solution will grant much higher level of the fluorescence than the well-known plasmonic fluorescent tag \cite{PEF}. Really, the number of molecules coupled to a plasmonic nanoparticle is much smaller that the number of molecules covering a rod with the thickness of few microns simply because it is much larger. Meanwhile, the mean Purcell factor of a typical plasmonic fluorescent tag is of the same order of magnitude as in our case (10-20). Here it is worth to note that the Purcell factor describes the gain in the fluorescence granted by a nanoantenna or a cavity and its mean value is equal to the mean value of the electric intensity enhancement \cite{Purcell}. 

The same observation concerns the surface-enhanced Raman scattering (SERS), where the molecules are conventionally coupled to plasmonic nanoparticles and it is difficult to properly locate them \cite{SERS}, whereas a substantial dielectric cylinder can be easily covered due to the self-assembly of long molecules to the dielectrics \cite{PEF}.
Moreover, for many biomedical analytes SERS without a dielectric spacer is forbidden, whereas the presence of the spacer critically decreases the 
local intensity enhancement \cite{SERS}. For these SERS schemes our cylinder promises a larger Raman gain.  

\subsection{Eigenmode pattern in the case of the diffraction-free incident beam}

\begin{figure}[ht]
\centering
\includegraphics[width=11cm]{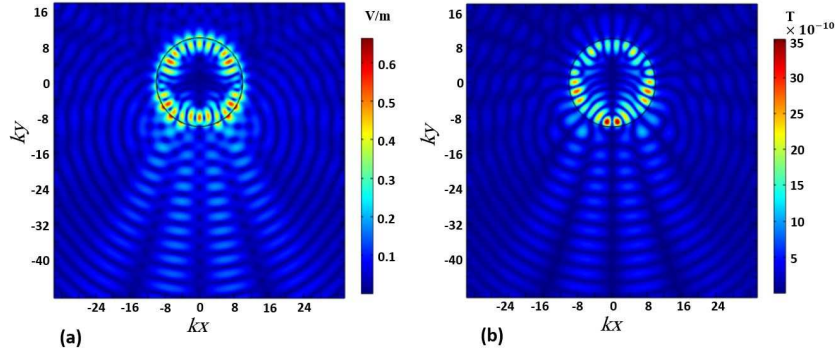}
\caption{Color map of the scattered electric field (instantaneous distribution of the magnitude) for the case $kR=10,\, n=1.7$ (a).
Color map of the scattered magnetic field for the same case (b). 
}
\label{Pic8}
\end{figure}

In this subsection we discuss why the electric field in our case turns out to be strongly enhanced outside a dielectric microparticle, whereas in the case of a usual wave beam or a plane wave incident on any resonant dielectric microparticle a so drastic enhancement occurs only inside it. 

Not only the internal field of the cylinder impinged by a wave beam comprises the cylindrical functions of high orders implying the small scale of spatial oscillations and, therefore, capable to grant the subwavelength field concentration due to their interference. The scattered field outside the cylinder comprises these functions as well. If we want to stress these terms, we may engineer the incident beam so that it would be nearly orthogonal to the large-scale oscillating terms and would maximally fit the small-scale ones. Our incident beam due to its symmetry does not excite the propagating waves in its paraxial region and, therefore, in this region the evanescent waves dominate. For small $\beta$ the whole cylinder is located in this paraxial region, and the excited mode pattern is less leaky and more evanescent than in the case of a single-wave excitation. 

We have studied our subwavelength effects for different values of $\beta$.  
For $\beta=0.01$ (the case reported in the present paper) and for $\beta=0.001$ all our results are practically the same. When $\beta=0.1$ the values of the intensity enhancement reported above decrease nearly twice and the focal spots enlarge twofold. When $\beta=0.2>\pi/2kR$ the near-field effects beyond the cylinder practically disappear. In this trivial case, each of two incident plane waves independently generates its own photonic nanojet tilted to the axis $y$ by angles $\pm \beta$.        

Now, let us inspect the electric field pattern depicted in  Fig.~\ref{Pic8}(a) for the case $\beta=0.01$. 
We see that the spatial maxima of the electric field corresponding to the resonant mode $m=13$ are extended in the radial directions and expand outside the cylinder. For a single plane wave incident along the same axis $y$ the field maxima of the mode $TM_{13}$ are nearly circular spots which do not expand outside. The magnetic field map depicted in Fig.~\ref{Pic8}(b) is not very different from that corresponding to the incidence of a single plane wave. Both shapes and sizes of the magnetic hot spots are visually the same.  The main difference is the split of two magnetic field maxima centered at the axis $y$ onto two spots with the exact zero between them because in the present case $H(\phi=0)=0$.  For the magnetic field outside the cylinder we have not find near-field effects. Only the magnitude and phase of the outer electric field feel the impact of the resonances.     

Thus, our beam of two plane waves with a small angle $2\beta$ between the wave vectors offers an interference effect of the resonant or nearly resonant eigenmodes induced in the microcylinder which drastically elongate the tails of the the modal maxima and increase the impact of the evanescent waves behind the cylinder -- on its surface and even at a small distance from it. 

\section{Conclusions}

In this work we have studied the incidence of a diffraction-free beam (cosine beam) on an optically thick dielectric cylinder.
In our case, the beam electric and magnetic intensities have deep minima on its optical axis and the cylinder is fully located in between two adjacent maxima. For this case we theoretically revealed two strong and unusual near-field effects 
which arise in free space in the vicinity of the cylinder. The first one is 
subwavelength focusing of the incident beam behind the cylinder. The second one is drastic enhancement of the electric intensity on the whole surface of the cylinder, especially high on its back side. In both cases, the evanescent wave fields dominate over the fields of propagating waves. 

The conversion of the incident propagating waves into evanescent ones in the area is so efficient, because the whole cylinder is located in the paraxial region of the incident beam, and in this region propagating spatial harmonics are mutually canceled in the scattered field. This cancellation is dictated by the symmetry of the incident beam.   

We believe that these theoretically revealed effects will be confirmed experimentally and think that they are promising for such applications, as label-free subwavelength imaging in the far field, cavity-enhanced fluorescence and cavity-enhanced Raman scattering. Probably, these effects will find even more nanophotonic applications, than we can suggest now. 

\section{Acknowledgement}

Funding by the Russian Foundation for the Basic Research (grant №18-02-00315) is acknowledged by V.K.
\section{Disclosures}
The authors declare no conflicts of interest.

\bibliography{paper}

\begin{thebibliography}{10}
\newcommand{\enquote}[1]{``#1''}

\bibitem{radial}
Q.-X. Zhu, \enquote{Description of the propagation of a radially polarized beam
  with the scalar kirchhoff diffraction,} {\protect\JournalTitle{Journal of
  Modern Optics}} \textbf{56}, 1621--1625 (2009).

\bibitem{cosine}
E.~Recami, M.~Zamboni-Rached, and L.~A. Ambrosio, \emph{Non-Diffracting Waves}
  (John Wiley Sons, Ltd, 2014).

\bibitem{Sergei}
A.~V. Osipov and S.~A. Tretyakov, \emph{Modern Electromagnetic Scattering
  Theory with Applications} (John Wiley Sons, Ltd, 2017).

\bibitem{PNJ}
B.~S. Luk'Yanchuk, Z.~B. Wang, W.~D. Song, and M.~H. Hong, \enquote{Particle on
  surface: 3d-effects in dry laser cleaning,} {\protect\JournalTitle{Applied
  Physics A}} \textbf{79}, 747--751 (2004).

\bibitem{Itagi}
A.~V. Itagi and W.~A. Challener, \enquote{Optics of photonic nanojets,}
  {\protect\JournalTitle{J. Opt. Soc. Am. A}} \textbf{22}, 2847--2858 (2005).

\bibitem{Taflove}
Z.~Chen, A.~Taflove, and V.~Backman, \enquote{Photonic nanojet enhancement of
  backscattering of light by nanoparticles: a potential novel visible-light
  ultramicroscopy technique,} {\protect\JournalTitle{Opt. Express}}
  \textbf{12}, 1214--1220 (2004).

\bibitem{Taflove1}
A.~Heifetz, S.-C. Kong, A.~V. Sahakian, A.~Taflove, and V.~Backman,
  \enquote{Photonic nanojets,} {\protect\JournalTitle{Journal of computational
  and theoretical nanoscience}} \textbf{6}, 1979--1992 (2009).

\bibitem{Hong}
Z.~Wang, W.~Guo, L.~Li, B.~Lukyanchuk, A.~Khan, Z.~Liu, Z.~Chen, and M.~Hong,
  \enquote{Optical virtual imaging at 50 nm lateral resolution with a
  white-light nanoscope,} {\protect\JournalTitle{Nature Communication}}
  \textbf{2}, 218 (2011).

\bibitem{Yang}
H.~Yang, R.~Trouillon, G.~Huszka, and M.~A.~M. Gijs, \enquote{Super-resolution
  imaging of a dielectric microsphere is governed by the waist of its photonic
  nanojet,} {\protect\JournalTitle{Nano Letters}} \textbf{16}, 4862--4870
  (2016).

\bibitem{Lecler}
S.~Lecler, S.~Perrin, A.~Leong-Hoi, and P.~Montgomery, \enquote{Photonic jet
  lens,} {\protect\JournalTitle{Scientific Reports}} \textbf{9}, 4725 (2019).

\bibitem{Zhou}
S.~Zhou, Y.~Deng, W.~Zhou, M.~Yu, H.~P. Urbach, and Y.~Wu, \enquote{Effects of
  whispering gallery mode in microsphere super-resolution imaging,}
  {\protect\JournalTitle{Applied Physics B}} \textbf{123}, 236 (2017).

\bibitem{Maslov}
A.~V. Maslov and V.~N. Astratov, \enquote{Optical nanoscopy with contact
  mie-particles: Resolution analysis,} {\protect\JournalTitle{Applied Physics
  Letters}} \textbf{110}, 261107 (2017).

\bibitem{Cang}
H.~Cang, A.~Salandrino, Y.~Wang, and X.~Zhang, \enquote{Adiabatic far-field
  sub-diffraction imaging,} {\protect\JournalTitle{Nature Communications}}
  \textbf{6}, 7942 (2015).

\bibitem{Astratov1}
A.~Maslov and V.~Astratov, \enquote{Resolution and reciprocity in
  microspherical nanoscopy: Point-spread function versus photonic nanojets,}
  {\protect\JournalTitle{Phys. Rev. Applied}} \textbf{11}, 064004 (2019).

\bibitem{Reza}
R.~Heydarian and K.~Simovski, \enquote{Role of the normal polarization in the
  far-field subwavelength imaging by a dielectric microsphere or
  microcylinder,} {\protect\JournalTitle{Journal of Optics}} \textbf{22},
  075002 (2020).

\bibitem{Klimov}
V.~Klimov, \enquote{Manifestation of extremely high-q pseudo-modes in
  scattering of a bessel light beam by a sphere,} {\protect\JournalTitle{Optics
  Letters}} \textbf{45}, 4300--4303 (2020).

\bibitem{PEF}
B.~Valeur and M.~N. Berberan-Santos, \emph{Molecular Fluorescence: Principles
  and Applications} (Wiley-VCH, 2012).

\bibitem{Purcell}
S.~Maslovski and C.~Simovski, \enquote{Purcell factor and local intensity
  enhancement in surface-enhanced raman scattering,}
  {\protect\JournalTitle{Nanophotonics}} \textbf{8}, 429--434 (2019).

\bibitem{SERS}
I.~Alessandri and J.~R. Lombardi, \enquote{Editorial: Surface enhanced raman
  scattering: New theoretical approaches, materials and strategies,}
  {\protect\JournalTitle{Frontiers in Chemistry}} \textbf{8}, 63 (2020).

\end{thebibliography}

\end{document}